\documentstyle[prl,aps]{revtex}

\begin{document}
\draft
\twocolumn
\author{
Yan V. Fyodorov$\S\P$\thanks{
On leave from Petersburg Nuclear Physics Institute, Gatchina
188350, Russia}}
\address{
$\S$Fachbereich Physik, Universit\"at-GH Essen,
D-45117 Essen, Germany
        }

\title{  Fluctuations in random $RL-C$ networks:
 non-linear $\sigma-$ model description}

\date{\today}
\maketitle
\begin{abstract}
 Disordered $RL-C$ networks are known to be an adequate model 
for describing fluctuations of 
electric fields in  a random metal-dielectric composite.
We show that under appropriate conditions 
the statistical properties of such a system can be 
studied in the framework of the Efetov's
non-linear $\sigma-$ model. 
This fact provides a direct link to the theory
of Anderson localization.

\end{abstract}

Optical properties of random metal-dielectric films (also
known as cermets or semicontinous metal films) 
attracted a lot of research interest recently,
both theoretically and experimentally, see \cite{KH,BBS,Entin,Luck}
and references therein. 
It was discovered that
for metal concentrations close to the percolation threshold 
the absorption of microwave radiation in such materials fluctuates
anomalously. In turn, these anomalous properties
 were traced back to high local field 
fluctuations detected in such compounds. 
A very insightful
approach to the problem\cite{KH,BBS,Entin,Luck} is to represent the system 
as a large random network
made of capacitors $C$ and inductances $L$, the latter being in
series with a weak resistor $R$. The network description naturally
arises when discretizing the equations satisfied by the scalar potential
of the electric field. The capacitors here are to model
dielectric bridges, whereas isolated metallic granulas are indeed
characterized by almost purely inductive response for frequencies $\omega$ 
of radiation such that $\omega_{\tau}\ll \omega \lesssim \omega_p$,
 with $\omega_p$ being the plasma frequency and $\omega_{\tau}$ being
the plasmon relaxation rate\cite{KH,BBS,Entin,Luck}.

For frequencies close to $\omega_0=1/(LC)^{1/2}$ an electromagnetic
response of such a network is dominated by resonance effects as long
as losses are small, i.e. the quality 
factor $Q=(L/C)^{1/2}R^{-1}$ is large.
The resonance frequencies can be determined as (generalized)
eigenvalues of some linear lattice operator arising when solving
the system of Kirchhoff equations $\sum_j \sigma_{ij}(v_i-v_j)=0$ for on-site potentials $v_i\equiv v({\bf
r_i})$\cite{Luck,my}. 
Here $\sigma_{ij}$ is the conductance between pair of nodes 
 ${\bf r}_i$ and ${\bf r}_j$ if two nodes are
connected by a direct bond and  $\sigma_{ij}=0$ otherwise.
 In the simplest case, 
one can think of the network being connected
to AC voltage by two external leads attached to lattice nodes with the 
coordinates ${\bf r}_A$ and ${\bf r}_B$, the
corresponding potentials being $v_A=e^{-i\omega t}$ and $v_B=0$, respectively. 
Omitting the common time-dependent
factor it is easy to see that the amplitude of the potential $v({\bf r_i})$
 at an internal lattice node ${\bf r}_i$ is given by:
\begin{eqnarray}
 v({\bf r_i})&=&\sum_{j}\left(\hat{D}^{-1}\right)_{ij}
\sigma_{Aj}\\ \nonumber  
\hat{D}_{ij}&=&\left(\sigma_{Ai}+\sigma_{Bi}+\
\sum_{j\ne i}\sigma_{ij}\right)\delta_{ij}-(1-\delta_{ij})\sigma_{ij}
\end{eqnarray}

In a random $RL-C$ network each 
nonzero conductance $\sigma_{ij}$ at frequency
$f=\omega/2\pi$ is equal to either $\sigma_0=iC\omega$
or $\sigma_1=(R+iL\omega)^{-1}$, with a specified probabilities
(in what follows we concentrate on the case of equal probability
for finding $L$ and $C$ bonds in the network).
Then it is convenient to introduce "symmetric" 
variables $h_{ij}$ such that
$h_{ij}=-1$ if $\sigma_{ij}=\sigma_0$ and $h_{ij}=1$ if
$\sigma_{ij}=\sigma_1$, so that $\sigma_{ij}=e_{ij}\frac{1}{2}
\left([\sigma_0+\sigma_1]+[\sigma_1-\sigma_0]h_{ij}\right)$,
with $e_{ij}=1$ for directly connected nodes and $e_{ij}=0$ 
otherwise. 
In terms of these variables we can write $D=\hat{H}-\lambda{\hat W}$,
where
\begin{eqnarray}
\hat{W}_{ij}&=&(Z+e_{Ai}+e_{Bi})\delta_{ij}-(1-\delta_{ij})e_{ij}\\
\hat{H}_{ij}&=&\left(\tilde{h}_{Ai}+\tilde{h}_{Bi}+\
\sum_{k\ne i}\tilde{h}_{ik}\right)\delta_{ij}-(1-\delta_{ij})\tilde{h}_{ij}
\end{eqnarray}
with $Z=\sum_j e_{ij}$ standing for the coordination number of the
lattice  and $\tilde{h}_{ij}=h_{ij}e_{ij}$. The frequency-dependent
parameter $\lambda$ is defined as 
\begin{equation}
\lambda=\frac{\sigma_0+\sigma_1}{\sigma_0-\sigma_1}\approx
\left(\frac{\omega}{\omega_0}-1\right)-\frac{i}{2Q}\equiv
\mbox{Re}\lambda-i\frac{\Gamma}{2}
\end{equation}
where we have made use of $\omega\approx \omega_0$ and $Q\gg 1$. 

We see, that statistics of the scalar potential $v({\bf r})$ (and hence of
the electric field ${\cal E}_{ij}$ proportional to the voltage difference
$v({\bf r}_i)-v({\bf r}_j)$ on the bond $ij$) 
is determined basically by properties of the operator $\hat{H}$.
The operators of such a type acting on a lattice were suggested to be 
called {\it Kirchhof Hamiltonians} (KH) in \cite{KH}. 
Off-diagonal entries of such a Hamiltonian
assume random values $\pm 1$ for directly connected nodes. This property
makes KH to be , in a sense, 
similar to a tight-binding Hamiltonian describing the
motion of a quantum particle on a disordered lattice with
an off-diagonal disorder. The latter model is a paradigmatic
one in the theory of Anderson localization. That kind of analogy 
first discussed in \cite{KH} led the authors to
relating the anomalous fluctuations of electric fields to localized 
properties of the corresponding eigenfunctions. Further numerical and
experimental work confirmed the qualitative validity of the suggested
picture.

At the same time, the question to which extent one 
can push forward the analogy between
the Anderson model and the Kirchhof Hamiltonian 
is far from being trivial. 
Indeed, the KH has a specific feature: the diagonal
entries $H_{ii}$ are strongly correlated with the off-diagonal ones
$H_{i\ne j}$. It is known that correlations of various kinds can
substantiallly modify the localisation behaviour, see e.g.\cite{cor}. 
Therefore, it is highly desirable to find an adequate approach allowing 
to shed more light on the question of equivalence between the models. 

The main goal of the paper is to show that the equivalence indeed
exists and the unifying concept is provided by the
so-called Efetov's supermatrix non-linear $\sigma-$ model (ENSM)\cite{Efbook}.
The latter model is known to be the most powerful tool in understanding the 
fluctuation phenomena in disordered conductors last decade, see e.g.
\cite{Sasha}.  
As a matter of fact, we derive ENSM from a version of the Kirchhof
Hamiltonian and thus provide a regular analytical
background for the quantitative description of statistical properties  
of the semicontinous films.

To derive ENSM from a microscopic random Hamiltonian one has to
exploit some large parameter which physically controls the
strength of the disorder. The experience of
 dealing with the usual tight binding models suggests that a role of 
such a parameter can be played, e.g. by a large radius of
connectivity $b$ (i.e. the
large coordination number $Z\sim b^d$)\cite{band,band1}. 
Formally, we consider a 
$d-$dimensional lattice of a linear size $L$ with unit 
lattice spacing and a connectivity radius  $b$. To facilitate bookkeeping 
of terms of the different order it is convenient  
to redefine $e_{ij}\to e_{ij} Z^{-1/2}$ where $e_{ij}=1$ for
$|{\bf r}_i-{\bf r}_j|\le b$ and $e_{ij}=0$ otherwise. 
The radius of connectivity $b$ is chosen to satisfy $1\ll b\ll L$.
Both inequalities
are important: $b\gg 1$ allows one to map the problem to ENSM , whereas
$b\ll L$ is necessary to ensure the adequate description of effects
of the Anderson localization. Indeed, as is shown recently\cite{my}
a full-connectivity $LC-$network with $b=L$ can be mapped on the {\it
zero-dimensional} version of the ENSM, which precludes the
localization effects to be taken into account.

To demonstrate the mapping it is instructive to address the simplest
nontrivial correlation function of the potentials:
$C_{i}(\Omega,\Gamma)=\langle v^*_{\omega_1}({\bf r}_{i})
v_{\omega_2}({\bf r}_{i})\rangle$
where we introduce the frequency difference
$\Omega\propto (\omega_1-\omega_2)/2\omega_0\ll 1$
and the brackets stand for the disorder averaging\cite{note}.

 Our starting expression is:
\begin{eqnarray}\label{cordef}
\nonumber
&&{\cal C}_{i}(\Omega,\Gamma)=
\left\langle\sum_{k_1,k_2}e_{Ak_1}e_{Ak_2}
\left(\tilde{h}_{Ak_1}-\lambda_1^*\right)
\left(\tilde{h}_{Ak_2}-\lambda_2\right)\right.\\
&& \times \left.
\left[\frac{1}{H-\lambda_1 W}\right]^*_{ik_1}
\left[\frac{1}{H-\lambda_2 W}\right]_{ik_2}\right\rangle
\end{eqnarray} 

To perform the disorder average we follow the standard procedure
and represent the matrix element of the 
resolvent $(H-\lambda W)^{-1}$ in terms of
the Gaussian integral:
\begin{eqnarray}
\nonumber
&&\left[\frac{1}{H-(\mbox{Re}\lambda\pm i\Gamma/2) W}\right]_{ik}
=\pm i\int \left[\prod_{l=1}^{N} d \Psi_{l}(\pm)\right] s_i^*(\pm)s_k(\pm)\\
&&\times \exp\{\pm\frac{i}{2}\sum_{m,n}^N\Psi_m^{\dagger}(\pm)
\left[W_{mn}(\mbox{Re}\lambda\pm i\Gamma/2)-H_{mn}\right]\Psi_n(\pm)\}
\end{eqnarray}
over 4-component supervectors $\Psi_{l}(\pm)$,
\begin{eqnarray}\label{suvec1}
\Psi_{l}(\pm)=
\left(
\begin{array}{c} 
\vec{S}_{l}(\pm)\\
\vec{\eta}_{l}(\pm)
\end{array}
\right),
\vec{S}_{l}(\pm)=
\left(
\begin{array}{c}
s_{l}(\pm)\\
s_{l}^{*}(\pm)
\end{array}
\right)\\ \nonumber
\vec{\eta}_{l}(\pm)=
\left(
\begin{array}{c}
\chi_{l}(\pm)\\ 
\chi_{l}^{*}(\pm)
\end{array}
\right),
d\Psi_l=\frac{ds_l ds_l^*}{2\pi}d\chi_l^*d\chi_l
\end{eqnarray}
with components $s_{l}(+),s_{l}(-);\quad l=1,2,...,N$ 
being complex commuting
variables and $\chi_{l}(+),\chi_{l}(-)$ forming the 
corresponding Grassmannian
parts of the supervectors $\Psi_{l}(\pm)$.

To facilitate the presentation, it is appropriate  
to anticipate few facts whose validity can be verified by the same method
as presented below. First of all, after averaging
the double sum in the expression Eq.(\ref{cordef}) is dominated 
in the limit $1\ll b \ll L$ by the diagonal terms with indices $k_1=k_2$.
Another fact which is useful to exploit from the very beginning
is that all the resonance frequencies are concentrated in an interval
of the order of $\delta \omega/\omega_0 \sim Z^{-1/2}$ around $\omega=\omega_0$,
so that the typical spacing $\Delta$ between the neighbouring
resonances is of the order of $\Delta \sim \omega_0/(NZ^{1/2})$, with 
$N\sim L^d$ being the total number of resonance frequencies.
We anticipate nontrivial correlations occuring on the frequency
scale $\omega_1-\omega_2\sim \Delta$\cite{my}. For this reason we scale
$\mbox{Re}\lambda_{1,2}=\left(r\pm \Omega/2N\right)/Z^{1/2}$, considering both
$r$ and $\Omega$ to be of the order of unity. By the same reasoning
we consider the losses to be small enough to ensure that 
$\gamma=\Gamma (NZ^{1/2})$ is of the order of unity. Physically this
requirement means that a typical resonance width $\omega_0 \Gamma$
is considered to be comparable with the typical resonance spacing $\Delta$.     

With these facts in mind, we can easily average 
the products of the resolvents  
 over $\tilde{h}_{ij}=\pm 1/Z^{1/2} e_{ij}$ in the limit $N\gg Z \gg 1$. 
All further steps follow the method used in 
\cite{my} (cf. \cite{band1,MF}) adopted to the present model.  
We have: 
\begin{eqnarray}\label{q1}
 &&{\cal C}_{i}(\Omega,\gamma)= 
\frac{1}{Z}\sum_{k}e_{Ak}\int \left[\prod_l d\Phi_l\right] 
s_i^*(-)s_i(+)s_k^*(+)s_k(-)\\ \nonumber
&&\times\exp\left\{\frac{i}{4NZ}(\Omega+i\gamma)\sum_{m<n}
e_{mn}\left(\Phi_m^{\dagger}-\Phi_n^{\dagger}\right)
\left(\Phi_m-\Phi_n\right)\right\} \\ \nonumber
&&\times \exp\left\{-
\frac{1}{16Z}\sum_{m,n}e_{mn}{\cal K}(\Phi_m,\Phi_n)\right\}
\end{eqnarray}
Here the integration goes over the $8-$component supervectors
$\Phi_l^{\dagger}=\left(\Psi_l^{\dagger}(+),\Psi_l^{\dagger}(-)\right)$
and
\begin{eqnarray}\label{kern}
&&{\cal K}(\Phi_a,\Phi_b)=\\ \nonumber
&&-4ir\left(\Phi_a^{\dagger}-\Phi_b^{\dagger}\right)
\hat{\Lambda}\left(\Phi_a-\Phi_b\right)
+\left[ \left(\Phi_a^{\dagger}-\Phi_b^{\dagger}\right)
\hat{\Lambda}\left(\Phi_a-\Phi_b\right)\right]^2
\end{eqnarray}
with $\hat{\Lambda}=\mbox{diag}(1,1,1,1,-1,-1,-1,-1)$. Only those terms are
 left in the exponent which will later on contribute to the 
final expressions in the discussed limit. 

Next step is to use the following functional 
Hubbard-Stratonovich transformation (cf.\cite{band1,MF,my}):
\begin{eqnarray} \nonumber
&&\exp\left\{-\frac{1}{16Z}\sum_{m,n}e_{mn}{\cal K}(\Phi_m,\Phi_n)\right\}=
\int{\cal D}(g) e^{\frac{i}{8}\sum_{m=1}^N g_m(\Phi_m)}\\ \nonumber
&&\times \exp\left\{-\frac{Z}{16}
\sum_{mn}\left[\hat{e}^{-1}\right]_{mn}\int d\Phi_ad\Phi_b
g_m(\Phi_a){\cal C}(\Phi_a,\Phi_b)g_n(\Phi_b)\right\}
\end{eqnarray}
where $\left[\hat{e}^{-1}\right]$ is the matrix inverse to the matrix
$\hat{e}=\left[e_{ij}\right]|_{i,j=1,....,N}$ and
the kernel ${\cal C}(\Phi_a,\Phi_b)$ is , in a sense, 
the inverse of a (symmetric) kernel ${\cal K}(\Phi_a,\Phi_b)$:
\begin{equation}\label{inv}
\int d\Phi {\cal K}(\Phi_a,\Phi){\cal
C}(\Phi,\Phi_b)=\delta(\Phi_a,\Phi_b) ,
\end{equation}
with $\delta(\Phi_a,\Phi_b)$ playing the role of a $\delta-$ functional
kernel in a space spanned by the functions $g(\Phi)$.

With the help of these relations one easily brings each term
of the sum  in Eq.(\ref{q1}) to the form:
\begin{equation}\label{funint}
\int {\cal D}(g) {\cal F}_{+,-}[g_i]{\cal F}_{-,+}[g_k]
\exp{\left[{\cal L}\{g\}+\delta{\cal L}_1\{g\}\right]}
\end{equation}
where
\begin{eqnarray}\label{zdec}
&&{\cal F}_{\pm,\mp}[g]=\frac{\int d\Phi\, s^*(\pm)s(\mp)
e^{\frac{i}{8}g(\Phi)}}{\int d\Phi\, e^{\frac{i}{8}g(\Phi)}}\\
&&{\cal L}\{g\}=\sum_m\ln{\int d\Phi e^{\frac{i}{8}g_m(\Phi)}}
\\ \nonumber
&&-\frac{Z}{16}\sum_{m,n}[e^{-1}]_{mn}\int d\Phi_ad\Phi_b
g_m(\Phi_a){\cal C}(\Phi_a,\Phi_b)g_n(\Phi_b)\quad,\\
&&\delta{\cal L}_1\{g\}=i\frac{(\Omega+i\gamma)}{4NZ} \sum_{m<n}
e_{mn} \\ \nonumber 
&&\times\frac{\int d\Phi_m d\Phi_n e^{\frac{i}{8}\left[
g_m(\Phi_m)+g_n(\Phi_n)\right]}
\left(\Phi_m^{\dagger}-\Phi_n^{\dagger}\right)\left(\Phi_m-\Phi_n\right)}
{\int d\Phi_m d\Phi_n\exp{\frac{i}{8}
\left[g_m(\Phi_m)+g_n(\Phi_n)\right]}} 
\end{eqnarray}
and we restricted ourselves by the leading order term in 
$\delta{\cal L}_1\{g\}$ which is only a small correction to ${\cal
L}\{g\}$. 

Next step is to evaluate the functional integral over $g(\Phi)$
by the saddle-point method, justified by two large parameters:
$Z$ and $N$. The saddle-point 
configuration $g^{(s)}_m(\Phi)$
can be found by requiring the vanishing variation of the "action"
${\cal L}\{g\}$ and satisfies the following system of equations:
\begin{equation}\label{fsp}
Z\sum_n[e^{-1}]_{mn}g^{(s)}_n(\Phi_a)=i\frac{\int d\Phi_b 
{\cal K}(\Phi_a,\Phi_b)e^{\frac{i}{8}g^{(s)}_m(\Phi_b)}}
{\int d\Phi_b e^{\frac{i}{8}g^{(s)}_m(\Phi_b)}}
\end{equation}
When deriving Eq.(\ref{fsp}) we have used Eq.(\ref{inv}).

Given the form of the kernel Eq.(\ref{kern}) and exploiting
$Z\sum_n [e^{-1}]_{mn}=1$ one can find 
 a space-independent solution $g^{(s)}_n(\Phi) \equiv g^{(s)}(\Phi)$
to equation Eq.(\ref{fsp}) :
\begin{equation}\label{ansatz}
g^{(s)}(\Phi_a)=4(r-G_1)(\Phi_a^{\dagger}\hat{\Lambda}\Phi_a)
+4iG_2(\Phi_a^{\dagger}\Phi_a)+i(\Phi_a^{\dagger}\hat{\Lambda}\Phi_a)^2
\end{equation}
provided the real coefficients $G_1,G_2$ are solutions of the system
of two conjugate equations:
\begin{eqnarray}\label{system}
G_2\pm iG_1&=&\int_0^{\infty}du
\exp\left\{\pm\frac{i}{2}u(r-G_1\pm iG_2)-\frac{u^2}{8}\right\}
\end{eqnarray}
Hints to verifying such a solution can be found in \cite{my}.

For further analysis it is very important that 
a solution to equations Eq.(\ref{system})
 exists for arbitrary $-\infty<r<\infty$
such that $G_2(r)>0$. Actually, the mean density of 
resonances is merely given by $\rho(r)=\frac{1}{\pi}G_2(r)$
\cite{my}. 

The most important consequence of the existence of the solution
$g^{(s)}(\Phi_a)$ in the form Eq.(\ref{ansatz}) with $G_2\ne 0$  is actually the
simultaneous existence of a whole {\it continous manifold} of the saddle points parametrized as:
\begin{equation}
g_T(\Phi_a)=g^{(s)}(\hat{T}\Phi_a),\quad \mbox{with} 
\quad \hat{T}^{\dagger}\hat{\Lambda}\hat{T}=\hat{\Lambda}
\end{equation}
 If it hadn't been
for  the condition $G_2\ne 0$ all these solutions would trivially 
coincide: $g_T(\Phi)\equiv g^{(s)}(\Phi)$ for
any $\hat{T}$ defined as above.
In the actual case  presence of the combination
$\Phi_a^{\dagger}\Phi_a$  which is {\it not invariant} with respect  
to a transformation $\Phi_a\to \hat{T}\Phi_a$ ensures the existence
of the mentioned manifold. This fact is just a particular 
manifestation of the phenomenon of spontaneous breakdown of symmetry.
 Different nontrivial solutions
are actually parametrized by the supermatrices $\hat{T}$ which are
elements of a graded coset space $UOSP(2,2/4)/UOSP(2/2)\otimes 
UOSP(2/2)$.

As a result, the functional integral over $g_m(\Phi)$
is dominated by "Goldstone modes" slowly changing in space 
 and parametrized as: 
$g^{(G)}_m(\Phi)=g^{(s)}(\hat{T}_m\Phi)$, with the matrices $\hat{T}_m$
which depend on the lattice site index $m=1,...,N$.
Our next step is to determine the 
effective action for these modes that is 
${\cal L}\{g^{(G)}_m\}+\delta{\cal L}_1\{g^{(G)}_m\}$.

First of all we notice that:
\[
\int d\Phi e^{\frac{i}{8}g_m^{(G)}(\Phi)}=1
\]
which allows one to perform the following manipulations:
\begin{eqnarray}
&&\int d\Phi_ad\Phi_b
g_m^{(G)}(\Phi_a){\cal C}(\Phi_a,\Phi_b)g_n^{(G)}(\Phi_b)\\ \nonumber
&&=i\int d\Phi g_n^{(G)}(\Phi)e^{\frac{i}{8}g_m^{(G)}(\Phi)}=
i\int d\Phi g^{(s)}(\hat{T}_n\hat{T}_m^{-1}\Phi)
e^{\frac{i}{8}g^{(s)}(\Phi)}\\ \nonumber
&&=-4G_2^2\mbox{Str}\left(\hat{T}_n\hat{T}^{-1}_m
\hat{\Lambda}\left(\hat{T}_n\hat{T}^{-1}_m\right)^{-1}\hat{\Lambda}\right)\\
\nonumber && = -4(\pi\rho)^2\mbox{Str}\left(\hat{T}^{-1}_m
\hat{\Lambda}\hat{T}_m\hat{T}^{-1}_n\hat{\Lambda}\hat{T}_n\right)
\end{eqnarray}
Here we first exploited the saddle-point equation Eq.(\ref{fsp}) together with 
Eq.(\ref{inv}) and then performed a change of variables: 
$\hat{T}_m\Phi\to \Phi$
which does not effect the measure $d\Phi$ because of the (pseudo)
unitarity of the matrices $\hat{T}$. Then the integral can be readily 
evaluated with help of the explicit form Eq.(\ref{ansatz}) (see e.g.
examples of similar calculations in \cite{my,MF}) 
and brought to the final form by
employing the cyclic permutation and the mentioned relation between $G_2$
and the density of resonances $\rho(r)$. 
In the very same way we also find:
\begin{eqnarray}
\nonumber &&\delta{\cal L}_1\{g\}=
\frac{(\Omega+i\gamma)}{4NZ} G_2 
\sum_{m<n}e_{mn}\mbox{Str}\left(\hat{T}^{-1}_m\hat{\Lambda}\hat{T}_m
\hat{\Lambda}+\hat{T}^{-1}_n\hat{\Lambda}\hat{T}_n\hat{\Lambda}\right)\\ 
&&=\pi\rho(\Omega+i\gamma)\frac{1}{4N}\sum_{m}
\mbox{Str}\left(\hat{T}^{-1}_m\hat{\Lambda}\hat{T}_m\hat{\Lambda}\right)
\end{eqnarray}
The pre-exponential factors ${\cal F}_{+,-}[g_i]$ are calculated 
analogously and are given by 
\[
{\cal F}_{+,-}[g_i]=\pi\rho\left[\hat{T}^{-1}_i
\hat{\Lambda}\hat{T}_i\right]^{2,5}\,;\,\,
{\cal F}_{-,+}[g_k]=\pi\rho\left[\hat{T}^{-1}_k\hat{\Lambda}
\hat{T}_k\right]^{6,1}
\]
where the indices of supermatrix elements are inherited from
the structure of the diadic product $\Phi\otimes \Phi^{\dagger}$.

We see, that calculating the correlation function Eq.(\ref{q1})
in the limit $L\gg b\gg 1$ amounts to evaluating the following integral over
the set of supermatrices $\hat{Q}_m=\hat{T}^{-1}_m\hat{\Lambda}\hat{T}_m$:
\begin{eqnarray}
&&{\cal C}_{i}=\frac{(\pi\rho)^2}{Z}\sum_k e_{Ak}
\int\left[\prod_{l=1}^N d\hat{Q}_l
\right] Q_i^{2,5}Q_k^{6,1}e^{{\cal L}(\hat{Q})}\\ \label{ac}
&&{\cal L}\left(\hat{Q}\right)=\frac{Z(\pi\rho)^2}{4}\sum_{m,n=1}^N
\left[e^{-1}\right]_{mn}
\mbox{Str} \hat{Q}_m\hat{Q}_n \\ \nonumber
&&+i\frac{\pi\rho(\Omega+i\gamma)}{4N}
\sum_{m=1}^N\mbox{Str}\left(\hat{Q}_m\hat{\Lambda}\right)
\end{eqnarray}

The action Eq.(\ref{ac}) is actually equivalent to
 the discretized version
of the supermatrix nonlinear $\sigma-$model well-studied in
the context of the Anderson localization \cite{Efbook,Sasha}.
Indeed, following \cite{band}
 notice that the condition $b\gg 1$ ensures slow variation of
the matrices $\hat{Q}_m$ with index $m$, so that it is legitimate
to pass from the lattice to continuum, when the action
assumes the standard form:
\begin{eqnarray}\label{stand}
{\cal L}\left(\hat{Q}\right)=\frac{\pi\rho}{8}\int d{\bf r} 
\mbox{Str}\left[D\left(\nabla\hat{Q}\right)^2
+2i\frac{(\Omega+i\gamma)}{V}\hat{Q}({\bf r})\hat{\Lambda}\right]\,,
\end{eqnarray}
where $D=\pi\rho Z^{-1}\sum_{|{\bf r}|<b}{\bf r}^2$
plays the role of the effective diffusion constant and $V=\int d{\bf r}$.
The large value of $D\propto b^2$ ensures that a typical spatial scale
$\xi$ of variation of $\hat{Q}({\bf r})$ is large: 
$\xi\propto b^2$. Thus, for
distances of the order of $b$ the matrices $\hat{Q}$ do not change
and therefore:
\begin{eqnarray}\label{fin}
{\cal C}_{i}=(\pi\rho)^2
\int{\cal D} \hat{Q}({\bf r}) 
Q^{2,5}({\bf r}_i)
Q^{6,1}({\bf r}_A)e^{{\cal L}(\hat{Q})}
\end{eqnarray} 

In conclusion, we managed to express the correlation function of the 
scalar potentials in terms of ENSM. Similar reduction is possible
also for other quantities of interest. 
As is well-known\cite{Efbook,Sasha}, explicit evaluation of the integrals
of the type Eq.(\ref{fin}) crucially depends on the parameter $g=2\pi\rho D
L^{d-2}$. For $g\to \infty$ the integral is dominated by the constant
configuration: $\hat{Q}({\bf r})=\hat{Q}_0$ and the result 
for Eq.(\ref{fin}) is very simple:
$\tilde{{\cal C}}_i=2i\pi\rho(r)/(\Omega+i\gamma)$. This is the so-called
"zero-dimensional" 
limit corresponding to the infinite-range connectivity model
\cite{my}. One can take into account weak 
localization effects finding $1/g$ corrections in any dimension
$d$, see e.g.\cite{Sasha}. For a quasi one-dimensional lattice
one can calculate integrals exactly in the limit $\Omega,\eta\to 0$
\cite{band}. One should be able also to study singular parts of
higher correlation functions. These questions are left for
further investigations.

The author is grateful to J.-M.Luck and V.M.Shalaev for their encouraging interest in the work. The present study was initiated during the 
author's stay at MPI for Complex Systems in Dresden
and supported by SFB-237 "Disorder and Large Fluctuations"
and by grant INTAS 97-1342.

\end{document}